\documentclass{article}
\usepackage{amsfonts,amssymb}

 \def\rql{\mbox{\strut\smash{\raisebox{-1.4ex}{''}}\hspace{-0.05em}}}
 \def\rqr{\mbox{\hspace{0em}``}}

\newcommand*{\mytheorem}{\mbox{T~h~e~o~r~e~m.~}}

\makeatletter
\renewcommand{\@oddfoot}{--- ~\today~ ---\hfil
\raisebox{0.3ex}{\tiny --- ~D.~A.~Arbatsky~ ``The certainty
principle''~ ---}\hfil --- ~\thepage~ ---}
 \makeatother

\begin{document}


\title{The certainty principle}
\author{D.~A.~Arbatsky\footnote{ http://daarb.narod.ru/ , http://wave.front.ru/ }}
\date{\today}
\maketitle

\begin{abstract}
The notion of the quantum angle is introduced. The quantum angle
turns out to be a metric on the set of physical states of a
quantum system. Its kinematics and dynamics is studied. The {\it
certainty}\/ principle for quantum systems is formulated and
proved. It turns out that the certainty principle is closely
connected with the Heisenberg uncertainty principle (it presents,
in some sense, an opposite point of view). But at the same time
the certainty principle allows to give rigorous formulations for
wider class of problems (it allows to rigorously interpret and
ground the analogous inequalities for the pairs of quantities like
time - energy, angle - angular momentum etc.)
\end{abstract}

\section*{The quantum angle and the certainty principle}

\paragraph{The notion of the quantum angle.}
As it is known, the set of states of any quantum system forms a
complex Hilbert space. We will denote it $\mathcal H$.

Elements of the space $\mathcal H$ we will denote as
$a,b\dots\in\mathcal H$.

The scalar product in $\mathcal H$ we will write as
 $\langle a|b\rangle$. It is linear with respect to the second
argument and anti-linear with respect to the first one.

The norm of a vector $a$ we will denote as
 $ \|a\| = \langle a | a \rangle ^{1/2}$.

Consider two non-zero vectors $a,b\in\mathcal H$. Let us define
between them {\it the quantum angle} by the formula:
\[
 \angle(a,b) = \arccos
     \frac{\big|\langle a|b\rangle\big|}{\|a\|\,\|b\|}  \ .
\]

According to Cauchy-Bunyakovsky-Schwarz inequality, under the
function $\arccos$ we have the value that is not greater than
unity. Therefore, the quantum angle is a real number:
\[
 \angle(a,b) \in \mathbb R \ , \qquad
 0 \leqslant \angle(a,b) \leqslant \frac\pi2 \ .
\]

For simplification of formulas we will later on always work with
normalized vectors: $\|a\|=1$, $\|b\|=1$. In this case the formula
for the angle is written simply as:
\[
 \angle(a,b) = \arccos \big|\langle a|b\rangle\big|  \ .
\]


\paragraph{Geometry of quantum angle.}
Let us consider the two extreme cases: $\angle(a,b)=0$ and
$\angle(a,b)=\pi/2$.

According to the Parseval equality, the first case takes place
when the vectors differ only by phase factor:
\[
 \angle (a,b) = 0 \qquad \Longleftrightarrow \qquad
 a\parallel b \ .
\]
From physical point of view, we can say that the corresponding
quantum states are {\it identical}.

The second case takes place when vectors are orthogonal:
\[
 \angle(a,b)=\frac\pi2 \qquad \Longleftrightarrow \qquad
 a \perp b \ .
\]
In this case we can say that the corresponding quantum states are
{\it completely different}.

Implying this physical terminology, which is used in the
considered extreme cases, let us introduce also the following
definition. Let us say that the states described by the vectors
$a$ and $b$ {\it differ not-substantially}, if $\angle(a,b)<1$;
let us also say that the states {\it differ substantially}, if
$\angle(a,b)\geqslant 1$.

\mytheorem{}
 {\it For any three vectors $a,b,c\in\mathcal H$ the inequality
takes place (``the triangle inequality''):
\[
 \angle(a,c) \leqslant \angle(a,b) + \angle(b,c) \ .
\]}

In order to prove this theorem, let us first notice that, so far
as the inequality is proved for three vectors, we can bound
ourself with the case when the Hilbert space $\mathcal H$ is three
dimensional: $\mathcal H=\mathbb C^3$.

Multiplying the vectors $a$, $b$, and $c$ by appropriate factors
and choosing orthonormal basis in $\mathcal H$ appropriately, we
can achieve that components of these vectors take the form:
\[
 a = (\begin{array}{ccc} 1   & 0   & 0   \end{array}) \ , \qquad
 b = (\begin{array}{ccc} b_1 & z   & b_3 \end{array}) \ , \qquad
 c = (\begin{array}{ccc} c_1 & c_2 & 0   \end{array}) \ ,
\]
where $c_1,c_2,b_1,b_3\in[0;1]$ are real non-negative numbers, and
$z\in\mathbb C$ is complex.

Let us introduce also the auxiliary vector $b'$:
\[
 b'= (\begin{array}{ccc} b_1 & |z| & b_3 \end{array}) \ .
\]

We have:
\[
 \angle(b',c) =
   \arccos\big(b_1c_1+|z|\,c_2\big) \, \leqslant
   \arccos|\,b_1c_1+zc_2\,| =
   \angle(b,c) \ .
\]

So far as the three vectors $a$, $b'$, and $c$ have real
coordinates and unit lengths, the triangle inequality for them is
the well known triangle inequality on the sphere in the real
three-dimensional Euclidean space:
\[
 \angle(a,c) \leqslant \angle(a,b') + \angle(b',c) \ .
\]

Combining the obtained inequalities we get:
\[
 \angle(a,c) \leqslant
 \angle(a,b') + \angle(b',c) =
 \angle(a,b) + \angle(b',c) \leqslant
 \angle(a,b) + \angle(b,c) \qquad \blacksquare
\]

Summarizing what was said above, we can say that the quantum angle
$\angle$, considered as a function on the set of pairs of physical
quantum states (considered to phase factor), is a {\it metric}.

\mytheorem{}
 {\it The metric space of physical quantum states with
the metric $\angle$ is complete.}

The proof of this theorem is not difficult, but it requires
substantial technical work. So we omit it here.


\paragraph{Kinematics of quantum angle.}
Let now the vector $r$ depend on the real parameter $t$:
$t\in\mathbb R$, $r(t)\in\mathcal H$, $\|\,r(t)\,\|=1$.

Let us define {\it the quantum velocity} $v(t)$ by the formula:
\[
 v(t) = \dot r(t) =
    \lim_{\delta t\to0}
    \frac{r(t+\delta t)-r(t)}{\delta t} \ .
\]

Let us define also {\it the quantum angular speed} $\omega(t)$ by
the formula:
\[
 \omega(t) = \lim_{\delta t\to0}
     \frac{\angle\,\big(\,r(t+\delta t)\,,\,r(t)\,\big)}{|\,\delta t\,|} \ .
\]

In order to express $\omega(t)$ through $v(t)$ let us decompose
$v(t)$ into the two orthogonal components:
\[
 v_\parallel(t) = r(t)\,\langle r(t)|v(t)\rangle \ , \qquad
 v_\perp(t) = v(t) - v_\parallel(t) \ .
\]

\mytheorem{}
 {\it The quantum angular speed is equal to the norm of the orthogonal
component of the quantum velocity:
\[
 \omega(t)=\|\,v_\perp(t)\,\| \ .
\]}

In order to prove this theorem let us use the Parseval equality to
change $\arccos$ to $\arcsin$:
\[
 \angle\,\big(\,r(t+\delta t)\,,\,r(t)\,\big) \,=\,
 \arccos\,\big|\,\langle\,r(t+\delta t)\,|\,r(t)\,\rangle\,\big| \,=\,
\]
\[
 \,=\,
 \arcsin\,
   \big\|\,
   r(t+\delta t) -
   r(t) \,\langle\, r(t) \,|\, r(t+\delta t) \,\rangle
   \,\big\| \,=\,
\]
\[
 \,=\,
   \arcsin\,
    \big\|\,
    r(t+\delta t) \,-\, r(t) \,-\,
    r(t) \,
     \langle\,
        r(t) \,|\, r(t+\delta t)-r(t)
     \,\rangle
    \,\big\|
 \,=\,
\]
\[
 \,=\,
   \arcsin\,
    \big\|\,
    v(t)\,\delta t \,+\, o(\delta t) \,-\,
    r(t) \,
     \langle\,
        r(t) \,|\, v(t)\,\delta t + o(\delta t)
     \,\rangle
    \,\big\|
 \,=\,
\]
\[
 \,=\,
   \arcsin\,
    \Big\|\,
     \big(v(t)\,-\,
     r(t) \,
     \langle\,
        r(t) \,|\, v(t)
     \,\rangle\big)
     \,\delta t \,+\, o(\delta t)
    \,\Big\|
 \,=\,
\]
\[
 \,=\,
   \arcsin\,
    \big\|\,
    v_\perp(t)\,\delta t \,+\, o(\delta t)
    \,\big\|
 \,=\,
   \arcsin\,
    \big(\,
    \|v_\perp(t)\|\,|\delta t| \,+\, o(\delta t)
    \,\big)
 \,=\,
\]
\[
 \,=\,
 \|v_\perp(t)\|\,|\delta t| \,+\, o(\delta t)  \qquad \blacksquare
\]

\mytheorem{}
 {\it The quantum angle satisfy the estimate:
\begin{equation}
 \angle\,\big(\, r(t_2) \,,\, r(t_1) \,\big)
 \ \leqslant\
 \left|\
 \int_{t_1}^{t_2} \omega(t) \, dt
 \ \right|
 \ \ .
 \label{AngleIneq}
\end{equation}}

For the proof let us use the triangle inequality:
\[
 \Big|\,
 \angle \big(\, r(t+\delta t) \,,\, r(t_1) \,\big) -
 \angle \big(\, r(t) \,,\, r(t_1) \,\big)
 \,\Big|
 \ \leqslant\
 \angle \big(\, r(t+\delta t) \,,\, r(t) \,\big)
 \ .
\]
Dividing by $|\delta t|$ and taking the limit $\delta t\to 0$, we
get:
\[
 \left|\
 \frac{d}{dt}\,\angle \big(\, r(t) \,,\, r(t_1) \,\big)
 \ \right|
 \ \leqslant\
 \omega(t)
 \ .
\]
Performing integration from $t_1$ to $t_2$, we get the desired
inequality. $\blacksquare$

In fact, the estimate~(\ref{AngleIneq}) is the best. Namely, there
is the following

\mytheorem{}
 {\it The quantum angle between two vectors $r_1$ and $r_2$ can be
expressed by the formula:
\[
 \angle\,(\, r_2 \,,\, r_1 \,)
 \ =\
 \min\ \left|\
 \int_{t_1}^{t_2} \omega(t) \, dt
 \ \right|
 \ \ ,
\]
where the minimum is taken among all curves $r(t)$ with ends in
$r_1$ and $r_2$: $r(t_1)=r_1$, $r(t_2)=r_2$.}

For the proof of the theorem let us twist the phase of $r_2$,
 $r_2\to r_2'=e^{i\alpha}r_2$, so that
 $\langle r_2'|r_1\rangle$ become real and
 $\langle r_2'|r_1\rangle\in[0,1]$.

Let us consider the linear shell of $r_2'$ and $r_1$,
 $\mathcal L(r_2',r_1)$. There we can choose an orthonormal basis
so that
\[
 r_1 = (\begin{array}{ccc} 1 & 0 \end{array}) \ , \qquad
 r_2'= (\begin{array}{ccc} a & b \end{array}) \ ,
\]
where $a,b\in[0,1]$ are real non-negative numbers.

Considering then $r_2'$ and $r_1$ as real vectors on Euclidean
plane we see that it is possible to stretch the circular arc
between them where the estimation integral is exactly equal to the
quantum angle. $\blacksquare$


\paragraph{Dynamics of quantum angle.}
Let us have now a strongly continuous one-parameter unitary group
$U(\delta s)=e^{-\,i\,\delta s\,A/\hbar}$, where $A=A^*$ is a
self-adjoint operator in the space of states $\mathcal H$ (it is
called the infinitesimal generator of $U(\delta s)$); $\delta
s\in\mathbb R$ is the parameter of the group;
 $\hbar\in\mathbb R$ is the Planck's constant.

And suppose now that the dependence of state vector on the
parameter $\delta s$ is defined by the formula:
\[
 r(\delta s) \,=\, |\,\delta s\,\rangle \,=\,
 U(\delta s)\,\rangle \,=\,
 e^{-\,i\,\delta s\,A/\hbar} \,\rangle \ .
\]
Here $|\,\delta s\,\rangle\in\mathcal H$ is another notation for
the state vector connected with parameter equal to $\delta s$;
$\rangle\in\mathcal H$ is a fixed ket-vector of state.

Let us suppose that the function $r(\delta s)$ is differentiable.
Then the quantum velocity is expressed by the formula:
\[
 \textstyle
 v(\delta s) \,=\,
 \frac1{i\hbar}\, A \, e^{-\,i\,\delta s\,A/\hbar} \,\rangle \,=\,
 \frac1{i\hbar}\, A \, |\,\delta s\,\rangle \ .
\]

The mean of the operator $A$ does not depend on time:
\[
 \overline A \,=\,
 \langle\,\delta s\,|\, A \, |\,\delta s\,\rangle \,=\,
 \langle\,e^{+\,i\,\delta s\,A/\hbar} \, A \,
    e^{-\,i\,\delta s\,A/\hbar} \,\rangle \,=\,
 \langle\, A \, e^{+\,i\,\delta s\,A/\hbar} \,
    e^{-\,i\,\delta s\,A/\hbar} \,\rangle \,=\,
 \langle\, A \, \rangle \ .
\]

Therefore the components of the quantum velocity can be written
just as:
\[
 \textstyle
 v_\parallel(\delta s) \,=\,
 |\,\delta s\,\rangle \, \langle\,\delta s\,|\,
 \frac1{i\hbar}\, A \, |\,\delta s\,\rangle \,=\,
 \frac1{i\hbar}\,\overline A \, |\,\delta s\,\rangle
\]
\[
 \textstyle
 v_\perp(\delta s) \,=\,
 \frac1{i\hbar}\,\left(A-\overline A\right) \,
 |\,\delta s\,\rangle \ .
\]

The quantum angular speed turns out to be independent of time
also:
\[
 \textstyle
 \omega(\delta s)
 \,=\,
 \|\,v_\perp(\delta s)\,\|
 \,=\,
 \frac1\hbar\,
 \langle\,\delta s\,|\,
    \left(A-\overline A\right)^2
 \, |\,\delta s\,\rangle^{1/2}
 \,=\,
 \frac1\hbar\,
 \langle\,
    e^{+\,i\,\delta s\,A/\hbar} \,
    \left(A-\overline A\right)^2 \,
    e^{-\,i\,\delta s\,A/\hbar}
 \,\rangle^{1/2}
 \,=\,
\]
\[
 \textstyle
 \,=\,
 \frac1\hbar\,
 \langle\,
    \left(A-\overline A\right)^2 \,
    e^{+\,i\,\delta s\,A/\hbar} \,
    e^{-\,i\,\delta s\,A/\hbar}
 \,\rangle^{1/2}
 \,=\,
 \frac1\hbar\,
 \langle\,
    \left(A-\overline A\right)^2
 \,\rangle^{1/2}
 \,=\,
 \frac1\hbar\, \Delta_\rangle A  \ .
\]
Here $\Delta_\rangle A$ is a short notation for the standard
deviation of $A$ in the state $\rangle$.

Consider now how the quantum angle $\angle(\, |\,\delta s\,\rangle
\,,\, \rangle \,)$ behaves in this case. Using for it the
estimate~(\ref{AngleIneq}) we have:
\[
 \angle(\, |\,\delta s\,\rangle \,,\, \rangle \,)
 \ \leqslant\
 \left|\
 \int_0^{\delta s} \omega(\sigma) \, d\sigma
 \ \right|
 \ =\
 \textstyle
 \frac1\hbar\,|\delta s|\,\Delta_\rangle A \ .
\]
From this inequality we obtain the

\mytheorem{}
 {\it So that under the action of strongly continuous one-parameter
unitary group $U(\delta s)=e^{-\,i\,\delta s\,A/\hbar}$ the
initial state vector $\rangle$ changes substantially, it is
necessary to satisfy the inequality:
\begin{equation}
 \begin{array}{|ccc|}
 \hline
 \phantom{1} & & \phantom{1}\\
  & |\delta s|\, \Delta_\rangle A \geqslant \hbar & \\
 \phantom{1} & & \phantom{1}\\
 \hline
 \end{array}
 \label{Certainty1}
\end{equation}}

By the example of the Schr\"odinger particle we will see that this
theorem turns out to be closely connected with the Heisenberg
uncertainty principle, but has other meaning. Taking into account
this connection, we can name this theorem {\it the certainty
principle}.

The inequality expressing the certainty principle can be written
also in the following way:
\begin{equation}
 \begin{array}{|ccc|}
 \hline
 \phantom{1} & & \phantom{1}\\
  & \Delta_\rangle (\delta s A) \geqslant \hbar & \\
 \phantom{1} & & \phantom{1}\\
 \hline
 \end{array}
 \label{Certainty2}
\end{equation}
In this form it can be naturally carried over to the case when
 $\delta s$ and $A$ are matrices.

\section*{Examples}

\paragraph{One-dimensional Schr\"odinger particle.}
 Consider the one-dimensional Schr\"odinger particle with the
coordinate defined by the variable $x$. Its state vector can be
written by the wave function $\psi(x)$. In its space of states the
group of shifts acts by the formula:
\[
 U(\delta x)\,\psi(x) = \psi(x-\delta x) \ .
\]
This group can be written in the form:
\[
 U(\delta x) = e^{-\,i\,\delta x\,P/\hbar} \ , \qquad
 P = -\,i\hbar\frac d{dx}   \ .
\]
Here $P$ is the operator of momentum.

Applying the certainty principle in the form~(\ref{Certainty1}),
we get:
\[
 |\delta x|\, \Delta_{\psi(x)} P \geqslant \hbar \ .
\]

If we take as $\psi(x)$ a well localized packet of de~Broglie
waves, that turns into zero outside of some interval $l$, then
from this inequality, in particular, we have that
 $l\geqslant\hbar/\Delta_{\psi(x)} P$: because when the packet
is moved to the distance $l$, the change of the quantum angle must
turn out to be greater than $1$ (namely, $\pi/2$).

So, the Heisenberg uncertainty principle, if it is understood in
qualitative sense, follows from the certainty principle.

But if we understand the Heisenberg uncertainty principle in
quantitative sense, according to the Pauli-Weyl inequality
\begin{equation}
 \Delta_{\psi(x)} X\,\Delta_{\psi(x)} P \,
 \geqslant\, \frac\hbar2 \ \ ,
 \label{PauliWeyl}
\end{equation}
then there is no direct connection between these two principles.

Furthermore, from physical point of view, the Heisenberg
uncertainty principle and the certainty principle are like two
points of view on the spread of the wave packet. The Heisenberg
uncertainty principle says, that the wave packet is spread,
because the {\it classical} state of the particle is {\it badly
defined}. On the other hand, the certainty principle states, that
the wave packet is spread, because the {\it quantum} state is {\it
well defined}.

To the three-dimensional case the certainty principle is easily
generalized in the form~(\ref{Certainty2}):
\[
 \Delta_{\psi(x)}\, (\,\delta x_i\, P_i\,) \geqslant \hbar \ ,
\]
where summation over $i$ is implied.


\paragraph{Schr\"odinger particle on circle.}
 Consider now a plane with Cartesian coordinates $x$ and $y$.
Let us have the circle $x^2+y^2=1$ defined on the plane. On this
circle as one-dimensional coordinate we can use the polar angle
$\varphi$, considered to $2\pi$.

The state vector of the Schr\"odinger particle on this circle can
be written by the wave function $\psi(\varphi)$. And
$\psi(\varphi+2\pi)=\psi(\varphi)$.

In the space of states the action of the rotation group is
naturally defined:
\[
 U(\delta \varphi)\,\psi(\varphi) = \psi(\varphi-\delta \varphi) \ .
\]
This group can be written in the form:
\[
 U(\delta\varphi) = e^{-\,i\,\delta\varphi\,J/\hbar} \ , \qquad
 J = -\,i\hbar\frac d{d\varphi}   \ .
\]
Here $J$ is the operator of angular momentum.

Using the certainty principle does not arise any difficulties:
\[
 |\delta\varphi|\, \Delta_{\psi(\varphi)} J \geqslant \hbar \ .
\]

As regards the uncertainty principle, a carrying over of the
inequality~(\ref{PauliWeyl}) to this case is
impossible\footnote{About some attempts that were made to suggest
an inequality like~(\ref{PauliWeyl}), see \cite{Dav1973}.}.

In the three-dimensional case the certainty principle also easily
gives:
\[
 \Delta_{\psi(x,y,z)}\, (\,\delta\varphi_i\, J_i\,)
 \geqslant \hbar \ .
\]


\paragraph{A system with Hamiltonian independent of time.}
 Let us have now a quantum system with a Hamiltonian $H$
independent of time. On the space of states we have the following
action of the group of time shifts:
\[
 U(\delta t)\,\rangle = e^{-\,i\,\delta t\,(-H)/\hbar}\rangle \ .
\]

The certainty principle in this case immediately gives:
\[
 |\delta t|\, \Delta_\rangle H \geqslant \hbar \ .
\]

If we apply this formula, for example, to estimation of the life
time of a quasi-stationary decaying state, then it states that its
typical life time is not less than Planck's constant divided by
the width of the corresponding energy level. And here all the
terms can be defined with exact mathematical sense.

As regards to attempts to formulate the Heisenberg uncertainty
principle for values time - energy, after Bohr has declared such a
principle (in qualitative sense) so many researches and
discussions were performed for clarification of its sense, that it
is possible to write about them a separate review. As far as I
know, a rigorous formulation of the uncertainty principle for this
case have not been formulated till now\footnote{See also the
discussion of this question by J.~Baez~\cite{Baez2000}.}.


\paragraph{Relativistic systems.}
 Consider now any relativistic quantum system.
On its space of states the Poincare group acts. As an example of
such a system any RCQ-quantized field can serve\footnote{All other
examples that I know either come from this or are mathematically
insolvent.} \cite{Arb2002,Arb2005wircq}. And let us restrict
ourself to the discussion of the case when the field turns out
quantized in the usual Hilbert space.

In this case the application of the certainty principle does not
meet any difficulties:
\[
 \textstyle
 \Delta_\rangle\,
   (\,-\,\delta x_\mu\, P_\mu\, +
      \frac12\,
      \delta\omega_{\mu\nu}\, J_{\mu\nu} \,) \geqslant \hbar \ ,
\]
where $P_\mu$ is the vector operator of energy-momentum,
$J_{\mu\nu}$ is the tensor operator of four-dimensional angular
momentum, $\delta x_\mu$ and $\delta\omega_{\nu\rho}$ are the
standard logarithmic coordinates of the Poincare group.

As regards to the application of the Heisenberg uncertainty
principle, it is unlikely to be possible. In the previous
paragraph we have seen that for the values time - energy it arose
great difficulties.

In this connection, even from the ideas relativistic invariance it
is clear that even for coordinates and momenta the situation
cannot be simple. And it turns out to be true, because it is known
that all attempts to introduce the notion of coordinates (as
self-adjoint operators on the space of states) for relativistic
systems are quite artificial\footnote{And it turns out that for
some systems introduction of good operators of coordinates is not
possible at all~\cite{NewtWig1949}.}.



\end{document}



\title{Принцип определённости}
\author{Д.~А.~Арбатский\footnote{ http://daarb.narod.ru/ , http://wave.front.ru/ }}
\date{\today}
\maketitle

\begin{abstract}
Вводится понятие квантового угла. Квантовый угол оказывается
метрикой на множестве физических состояний квантовой системы.
Изучается его кинематика и динамика. Формулируется и доказывается
принцип {\it определённости}\/ для квантовых систем. Принцип
определённости оказывается тесно связанным с принципом
неопределённости Гейзенберга (представляет, в определённом смысле,
обратную точку зрения). При этом принцип определённости позволяет
дать строгие формулировки для более широкого класса задач (удаётся
строго истолковать и обосновать аналогичные неравенства для пар
величин типа время - энергия, угол - момент импульса и т. п.)
\end{abstract}

\section*{Квантовый угол и принцип определённости}

\paragraph{Понятие квантового угла.}
Как известно, множество состояний любой квантовой системы образует
комплексное гильбертово пространство. Мы будем обозначать его
 $\mathcal H$.

Элементы пространства $\mathcal H$ будем обозначать как
$a,b\dots\in\mathcal H$.

Скалярное произведение в $\mathcal H$ будем записывать как
 $\langle a|b\rangle$. Оно линейно по второму аргументу и антилинейно
по первому.

Норму вектора $a$ будем обозначать как
 $ \|a\| = \langle a | a \rangle ^{1/2}$.

Рассмотрим два ненулевых вектора $a,b\in\mathcal H$. Определим
между ними {\it квантовый угол} формулой:
\[
 \angle(a,b) = \arccos
     \frac{\big|\langle a|b\rangle\big|}{\|a\|\,\|b\|}  \ .
\]

В соответствии с неравенством Коши-Буняковского-Шварца, под знаком
$\arccos$ стоит величина, не превосходящая единицы. Следовательно,
квантовый угол является вещественным числом:
\[
 \angle(a,b) \in \mathbb R \ , \qquad
 0 \leqslant \angle(a,b) \leqslant \frac\pi2 \ .
\]

Для упрощения формул далее мы будем всегда работать с
нормированными векторами: $\|a\|=1$, $\|b\|=1$. В этом случае
формула для угла записывается просто как:
\[
 \angle(a,b) = \arccos \big|\langle a|b\rangle\big|  \ .
\]


\paragraph{Геометрия квантового угла.}
Рассмотрим два крайних случая: $\angle(a,b)=0$ и
$\angle(a,b)=\pi/2$.

В соответствии с равенством Парсеваля, первый случай имеет место,
когда векторы отличаются лишь фазовым множителем:
\[
 \angle (a,b) = 0 \qquad \Longleftrightarrow \qquad
 a\parallel b \ .
\]
С физической точки зрения, можно сказать, что соответствующие
квантовые состояния {\it одинаковы}.

Второй случай имеет место, когда векторы ортогональны:
\[
 \angle(a,b)=\frac\pi2 \qquad \Longleftrightarrow \qquad
 a \perp b \ .
\]
В этом случае можно сказать, что соответствующие квантовые
состояния {\it полностью различны}.

Имея в виду эту физическую терминологию, употребляемую в
рассмотренных крайних случаях, введём также следующее определение.
Будем говорить, что состояния, описываемые векторами $a$ и $b$,
{\it отличаются несущественно}, если $\angle(a,b)<1$; также будем
говорить, что состояния {\it отличаются существенно}, если
$\angle(a,b)\geqslant 1$.

\mytheorem{}
 {\it Для любых трёх векторов $a,b,c\in\mathcal H$ имеет место
неравенство (\rql неравенство тре\-у\-голь\-ни\-ка\rqr):
\[
 \angle(a,c) \leqslant \angle(a,b) + \angle(b,c) \ .
\]}

Чтобы доказать эту теорему, заметим сначала, что, поскольку
неравенство доказывается для трёх векторов, можно ограничиться
случаем, когда гильбертово пространство $\mathcal H$ трёхмерно:
 $\mathcal H=\mathbb C^3$.

Домножая векторы $a$, $b$ и $c$ на подходящие множители и выбирая
ортонормированный базис в $\mathcal H$ надлежащим образом, можно
добиться того, чтобы компоненты этих векторов приняли вид:
\[
 a = (\begin{array}{ccc} 1   & 0   & 0   \end{array}) \ , \qquad
 b = (\begin{array}{ccc} b_1 & z   & b_3 \end{array}) \ , \qquad
 c = (\begin{array}{ccc} c_1 & c_2 & 0   \end{array}) \ ,
\]
где $c_1,c_2,b_1,b_3\in[0;1]$~--- вещественные неотрицательные
числа, а $z\in\mathbb C$ комплексно.

Введём также вспомогательный вектор $b'$:
\[
 b'= (\begin{array}{ccc} b_1 & |z| & b_3 \end{array}) \ .
\]

Имеем:
\[
 \angle(b',c) =
   \arccos\big(b_1c_1+|z|\,c_2\big) \, \leqslant
   \arccos|\,b_1c_1+zc_2\,| =
   \angle(b,c) \ .
\]

Поскольку три вектора $a$, $b'$ и $c$ имеют вещественные
координаты и единичные длины, для них неравенство треугольника~---
это хорошо известное неравенство треугольника на сфере в
вещественном трёхмерном евклидовом пространстве:
\[
 \angle(a,c) \leqslant \angle(a,b') + \angle(b',c) \ .
\]

Комбинируя приведённые неравенства, получаем:
\[
 \angle(a,c) \leqslant
 \angle(a,b') + \angle(b',c) =
 \angle(a,b) + \angle(b',c) \leqslant
 \angle(a,b) + \angle(b,c) \qquad \blacksquare
\]

Резюмируя сказанное выше, можно сказать, что квантовый угол
$\angle$, рассматриваемый как функция на множестве пар физических
квантовых состояний (рассматриваемых с точностью до фазового
множителя), является {\it метрикой}.

\mytheorem{}
 {\it Метрическое пространство физических квантовых состояний с
метрикой $\angle$ полно.}

Доказательство этой теоремы не сложно, но требует значительной
технической работы. Поэтому мы его здесь опустим.


\paragraph{Кинематика квантового угла.}
Пусть теперь вектор $r$ зависит от вещественного параметра $t$:
$t\in\mathbb R$, $r(t)\in\mathcal H$, $\|\,r(t)\,\|=1$.

Определим {\it квантовую скорость} $v(t)$ равенством:
\[
 v(t) = \dot r(t) =
    \lim_{\delta t\to0}
    \frac{r(t+\delta t)-r(t)}{\delta t} \ .
\]

Определим также {\it квантовую угловую скорость} $\omega(t)$
равенством:
\[
 \omega(t) = \lim_{\delta t\to0}
     \frac{\angle\,\big(\,r(t+\delta t)\,,\,r(t)\,\big)}{|\,\delta t\,|} \ .
\]

Чтобы выразить $\omega(t)$ через $v(t)$ разложим $v(t)$ на две
ортогональные составляющие:
\[
 v_\parallel(t) = r(t)\,\langle r(t)|v(t)\rangle \ , \qquad
 v_\perp(t) = v(t) - v_\parallel(t) \ .
\]

\mytheorem{}
 {\it Квантовая угловая скорость равна норме ортогональной
составляющей квантовой скорости:
\[
 \omega(t)=\|\,v_\perp(t)\,\| \ .
\]}

Чтобы доказать эту теорему, используем равенство Парсеваля для
замены $\arccos$ на $\arcsin$:
\[
 \angle\,\big(\,r(t+\delta t)\,,\,r(t)\,\big) \,=\,
 \arccos\,\big|\,\langle\,r(t+\delta t)\,|\,r(t)\,\rangle\,\big| \,=\,
\]
\[
 \,=\,
 \arcsin\,
   \big\|\,
   r(t+\delta t) -
   r(t) \,\langle\, r(t) \,|\, r(t+\delta t) \,\rangle
   \,\big\| \,=\,
\]
\[
 \,=\,
   \arcsin\,
    \big\|\,
    r(t+\delta t) \,-\, r(t) \,-\,
    r(t) \,
     \langle\,
        r(t) \,|\, r(t+\delta t)-r(t)
     \,\rangle
    \,\big\|
 \,=\,
\]
\[
 \,=\,
   \arcsin\,
    \big\|\,
    v(t)\,\delta t \,+\, o(\delta t) \,-\,
    r(t) \,
     \langle\,
        r(t) \,|\, v(t)\,\delta t + o(\delta t)
     \,\rangle
    \,\big\|
 \,=\,
\]
\[
 \,=\,
   \arcsin\,
    \Big\|\,
     \big(v(t)\,-\,
     r(t) \,
     \langle\,
        r(t) \,|\, v(t)
     \,\rangle\big)
     \,\delta t \,+\, o(\delta t)
    \,\Big\|
 \,=\,
\]
\[
 \,=\,
   \arcsin\,
    \big\|\,
    v_\perp(t)\,\delta t \,+\, o(\delta t)
    \,\big\|
 \,=\,
   \arcsin\,
    \big(\,
    \|v_\perp(t)\|\,|\delta t| \,+\, o(\delta t)
    \,\big)
 \,=\,
\]
\[
 \,=\,
 \|v_\perp(t)\|\,|\delta t| \,+\, o(\delta t)  \qquad \blacksquare
\]

\mytheorem{}
 {\it Для квантового угла справедлива оценка:
\begin{equation}
 \angle\,\big(\, r(t_2) \,,\, r(t_1) \,\big)
 \ \leqslant\
 \left|\
 \int_{t_1}^{t_2} \omega(t) \, dt
 \ \right|
 \ \ .
 \label{AngleIneq}
\end{equation}}

Для доказательства воспользуемся неравенством треугольника:
\[
 \Big|\,
 \angle \big(\, r(t+\delta t) \,,\, r(t_1) \,\big) -
 \angle \big(\, r(t) \,,\, r(t_1) \,\big)
 \,\Big|
 \ \leqslant\
 \angle \big(\, r(t+\delta t) \,,\, r(t) \,\big)
 \ .
\]
Поделив на $|\delta t|$ и переходя к пределу $\delta t\to 0$,
получаем:
\[
 \left|\
 \frac{d}{dt}\,\angle \big(\, r(t) \,,\, r(t_1) \,\big)
 \ \right|
 \ \leqslant\
 \omega(t)
 \ .
\]
Интегрируя от $t_1$ до $t_2$, получаем доказываемое неравенство.
$\blacksquare$

В действительности, оценка~(\ref{AngleIneq}) является наилучшей.
Именно, справедлива следующая

\mytheorem{}
 {\it Квантовый угол между двумя векторами $r_1$ и $r_2$ может быть задан
формулой:
\[
 \angle\,(\, r_2 \,,\, r_1 \,)
 \ =\
 \min\ \left|\
 \int_{t_1}^{t_2} \omega(t) \, dt
 \ \right|
 \ \ ,
\]
где минимум берётся среди всех кривых $r(t)$ с концами в $r_1$ и
$r_2$: $r(t_1)=r_1$, $r(t_2)=r_2$.}

Для доказательства теоремы подкрутим фазу $r_2$,
 $r_2\to r_2'=e^{i\alpha}r_2$, так, чтобы
 $\langle r_2'|r_1\rangle$ было вещественно и
 $\langle r_2'|r_1\rangle\in[0,1]$.

Далее, рассмотрим линейную оболочку $r_2'$ и $r_1$,
 $\mathcal L(r_2',r_1)$. В ней можно выбрать ортонормированный
базис так, чтобы
\[
 r_1 = (\begin{array}{ccc} 1 & 0 \end{array}) \ , \qquad
 r_2'= (\begin{array}{ccc} a & b \end{array}) \ ,
\]
где $a,b\in[0,1]$ --- вещественные неотрицательные числа.

Рассматривая далее $r_2'$ и $r_1$ как вещественные векторы на
евклидовой плоскости, замечаем, что между ними можно провести
дугу, на которой оценочный интеграл как раз равняется квантовому
углу. $\blacksquare$


\paragraph{Динамика квантового угла.}
Пусть теперь мы имеем сильно непрерывную однопараметрическую
унитарную группу $U(\delta s)=e^{-\,i\,\delta s\,A/\hbar}$, где
$A=A^*$~--- самосопряжённый оператор в пространстве состояний
$\mathcal H$ (он называется инфинитезимальным генератором
$U(\delta s)$); $\delta s\in\mathbb R$~--- параметр группы;
$\hbar\in\mathbb R$~--- постоянная Планка.

И пусть теперь зависимость вектора состояния от параметра
 $\delta s$ задаётся формулой:
\[
 r(\delta s) \,=\, |\,\delta s\,\rangle \,=\,
 U(\delta s)\,\rangle \,=\,
 e^{-\,i\,\delta s\,A/\hbar} \,\rangle \ .
\]
Здесь $|\,\delta s\,\rangle\in\mathcal H$~--- другое обозначение
вектора состояния при параметре, равном $\delta s$;
$\rangle\in\mathcal H$~--- фиксированный кет-вектор состояния.

Будем полагать, что функция $r(\delta s)$ дифференцируема. Тогда
квантовая скорость выражается формулой:
\[
 \textstyle
 v(\delta s) \,=\,
 \frac1{i\hbar}\, A \, e^{-\,i\,\delta s\,A/\hbar} \,\rangle \,=\,
 \frac1{i\hbar}\, A \, |\,\delta s\,\rangle \ .
\]

Среднее значение оператора $A$ не зависит от времени:
\[
 \overline A \,=\,
 \langle\,\delta s\,|\, A \, |\,\delta s\,\rangle \,=\,
 \langle\,e^{+\,i\,\delta s\,A/\hbar} \, A \,
    e^{-\,i\,\delta s\,A/\hbar} \,\rangle \,=\,
 \langle\, A \, e^{+\,i\,\delta s\,A/\hbar} \,
    e^{-\,i\,\delta s\,A/\hbar} \,\rangle \,=\,
 \langle\, A \, \rangle \ .
\]

Поэтому составляющие квантовой скорости могут быть просто записаны
как:
\[
 \textstyle
 v_\parallel(\delta s) \,=\,
 |\,\delta s\,\rangle \, \langle\,\delta s\,|\,
 \frac1{i\hbar}\, A \, |\,\delta s\,\rangle \,=\,
 \frac1{i\hbar}\,\overline A \, |\,\delta s\,\rangle
\]
\[
 \textstyle
 v_\perp(\delta s) \,=\,
 \frac1{i\hbar}\,\left(A-\overline A\right) \,
 |\,\delta s\,\rangle \ .
\]

Квантовая угловая скорость оказывается также независящей от
времени:
\[
 \textstyle
 \omega(\delta s)
 \,=\,
 \|\,v_\perp(\delta s)\,\|
 \,=\,
 \frac1\hbar\,
 \langle\,\delta s\,|\,
    \left(A-\overline A\right)^2
 \, |\,\delta s\,\rangle^{1/2}
 \,=\,
 \frac1\hbar\,
 \langle\,
    e^{+\,i\,\delta s\,A/\hbar} \,
    \left(A-\overline A\right)^2 \,
    e^{-\,i\,\delta s\,A/\hbar}
 \,\rangle^{1/2}
 \,=\,
\]
\[
 \textstyle
 \,=\,
 \frac1\hbar\,
 \langle\,
    \left(A-\overline A\right)^2 \,
    e^{+\,i\,\delta s\,A/\hbar} \,
    e^{-\,i\,\delta s\,A/\hbar}
 \,\rangle^{1/2}
 \,=\,
 \frac1\hbar\,
 \langle\,
    \left(A-\overline A\right)^2
 \,\rangle^{1/2}
 \,=\,
 \frac1\hbar\, \Delta_\rangle A  \ .
\]
Здесь $\Delta_\rangle A$ является сокращённым обозначением для
среднеквадратичного отклонения $A$ в состоянии $\rangle$.

Рассмотрим теперь, как ведёт себя в данном случае квантовый угол
$\angle(\, |\,\delta s\,\rangle \,,\, \rangle \,)$. Применяя для
него оценку~(\ref{AngleIneq}), имеем:
\[
 \angle(\, |\,\delta s\,\rangle \,,\, \rangle \,)
 \ \leqslant\
 \left|\
 \int_0^{\delta s} \omega(\sigma) \, d\sigma
 \ \right|
 \ =\
 \textstyle
 \frac1\hbar\,|\delta s|\,\Delta_\rangle A \ .
\]
Из этого неравенства вытекает

\mytheorem{}
 {\it Чтобы под действием сильно непрерывной однопараметрической унитарной
группы $U(\delta s)=e^{-\,i\,\delta s\,A/\hbar}$ начальный вектор
состояния $\rangle$ изменился существенно необходимо, чтобы
выполнялось неравенство:
\begin{equation}
 \begin{array}{|ccc|}
 \hline
 \phantom{1} & & \phantom{1}\\
  & |\delta s|\, \Delta_\rangle A \geqslant \hbar & \\
 \phantom{1} & & \phantom{1}\\
 \hline
 \end{array}
 \label{Certainty1}
\end{equation}}

На примере шрёдингеровской частицы мы увидим, что эта теорема
оказывается тесно связанной с принципом неопределённости
Гейзенберга, но имеет иной смысл. Имея в виду эту связь, можно
назвать эту теорему {\it принципом определённости}.

Неравенство, выражающее принцип определённости, можно записать ещё
следующим образом:
\begin{equation}
 \begin{array}{|ccc|}
 \hline
 \phantom{1} & & \phantom{1}\\
  & \Delta_\rangle (\delta s A) \geqslant \hbar & \\
 \phantom{1} & & \phantom{1}\\
 \hline
 \end{array}
 \label{Certainty2}
\end{equation}
В таком виде оно естественно переносится на случай, когда
 $\delta s$ и $A$ являются матрицами.

\section*{Примеры}

\paragraph{Одномерная шрёдингеровская частица.}
 Рассмотрим одномерную шрёдингеровскую частицу, координата которой
задаётся переменной $x$. Её вектор состояния может быть записан с
помощью волновой функции $\psi(x)$. В её пространстве состояний
действует группа сдвигов по формуле:
\[
 U(\delta x)\,\psi(x) = \psi(x-\delta x) \ .
\]
Эта группа может быть записана в виде:
\[
 U(\delta x) = e^{-\,i\,\delta x\,P/\hbar} \ , \qquad
 P = -\,i\hbar\frac d{dx}   \ .
\]
При этом $P$ является оператором импульса.

Применяя принцип определённости в форме~(\ref{Certainty1}),
получаем:
\[
 |\delta x|\, \Delta_{\psi(x)} P \geqslant \hbar \ .
\]

Если в качестве $\psi(x)$ взять хорошо локализованный пакет
дебройлевских волн, обращающийся в нуль вне некоторого интервала
$l$, то из этого неравенства, в частности, вытекает, что
 $l\geqslant\hbar/\Delta_{\psi(x)} P$: ведь при смещении пакета на
расстояние $l$ изменение квантового угла должно оказаться больше
$1$ (а именно~--- $\pi/2$).

Таким образом, принцип неопределённости Гейзенберга, понимаемый в
качественном смысле, вытекает из принципа определённости.

Если же понимать принцип неопределённости Гейзенберга в
количественном смысле, в соответствии с неравенством Паули-Вейля
\begin{equation}
 \Delta_{\psi(x)} X\,\Delta_{\psi(x)} P \,
 \geqslant\, \frac\hbar2 \ \ ,
 \label{PauliWeyl}
\end{equation}
то непосредственной математической связи между этими принципами
нет.

При этом, с физической точки зрения, принцип неопределённости
Гейзенберга и принцип определённости являются как бы двумя точками
зрения на размытость волнового пакета. Принцип неопределённости
Гейзенберга говорит, что волновой пакет является протяжённым,
поскольку {\it классическое} состояние частицы является {\it плохо
определённым}. Принцип же определённости утверждает, что волновой
пакет является протяжённым, поскольку {\it квантовое} состояние
является {\it хорошо определённым}.

На трёхмерный случай принцип определённости без труда обобщается в
форме~(\ref{Certainty2}):
\[
 \Delta_{\psi(x)}\, (\,\delta x_i\, P_i\,) \geqslant \hbar \ ,
\]
где по $i$ предполагается суммирование.


\paragraph{Шрёдингеровская частица на окружности.}
 Рассмотрим теперь плоскость с декартовыми координатами $x$ и $y$.
Пусть на этой плоскости задана окружность $x^2+y^2=1$. На этой
окружности в качестве одномерной координаты можно использовать
полярный угол $\varphi$, рассматриваемый с точностью до $2\pi$.

Вектор состояния шрёдингеровской частицы на этой окружности может
быть записан с помощью волновой функции $\psi(\varphi)$. При этом
$\psi(\varphi+2\pi)=\psi(\varphi)$.

В пространстве состояний при этом естественно определено действие
группы вращений:
\[
 U(\delta \varphi)\,\psi(\varphi) = \psi(\varphi-\delta \varphi) \ .
\]
Эта группа может быть записана в виде:
\[
 U(\delta\varphi) = e^{-\,i\,\delta\varphi\,J/\hbar} \ , \qquad
 J = -\,i\hbar\frac d{d\varphi}   \ .
\]
При этом $J$ является оператором момента импульса.

Применение принципа определённости не вызывает каких-либо
трудностей:
\[
 |\delta\varphi|\, \Delta_{\psi(\varphi)} J \geqslant \hbar \ .
\]

Что же касается применения принципа неопределённости, то
перенесение неравенства~(\ref{PauliWeyl}) на этот случай
невозможно\footnote{О предпринимавшихся попытках предложить
неравенство, аналогичное~(\ref{PauliWeyl}), см. \cite{Dav1973}.}.

В трёхмерном случае принцип определённости также легко даёт:
\[
 \Delta_{\psi(x,y,z)}\, (\,\delta\varphi_i\, J_i\,)
 \geqslant \hbar \ .
\]


\paragraph{Система с гамильтонианом, не зависящим от времени.}
 Пусть теперь имеется квантовая система с гамильтонианом $H$, не
зависящим от времени. При этом на пространстве состояний
оказывается определённым действие группы временных сдвигов:
\[
 U(\delta t)\,\rangle = e^{-\,i\,\delta t\,(-H)/\hbar}\rangle \ .
\]

Принцип определённости в этом случае сразу даёт:
\[
 |\delta t|\, \Delta_\rangle H \geqslant \hbar \ .
\]

Если эту формулу применить, например, к оценке времени жизни
квазистационарного распадающегося состояния, то она утверждает,
что характерное время жизни у него не менее чем постоянная Планка,
поделённая на ширину соответствующего уровня энергии. При этом
всем терминам можно придать точный математический смысл.

Что же касается попыток сформулировать принцип неопределённости
Гейзенберга для величин время - энергия, то с того времени, как
Бор провозгласил такой принцип (в качественном смысле), для
уточнения его смысла было проведено столько исследований и
дискуссий, что по ним можно было бы написать отдельный обзор.
Насколько мне известно, строгой формулировки принципа
неопределённости для этого случая получить так и не
удалось\footnote{См. также обсуждение этого вопроса
Дж.~Баезом~\cite{Baez2000}.}.


\paragraph{Релятивистские системы.}
 Рассмотрим теперь произвольную релятивистскую квантовую систему.
На её пространстве состояний действует группа Пуанкаре. Примером
такой системы может служить любое РКК-квантованное
поле\footnote{Все другие известные мне примеры либо вытекают из
указанного, либо математически несостоятельны.}
\cite{Arb2002,Arb2005wircq}. И ограничимся здесь обсуждением
случая, когда поле оказывается квантованным в обычном гильбертовом
пространстве.

В этом случае применение принципа определённости не сталкивается с
какими-либо трудностями:
\[
 \textstyle
 \Delta_\rangle\,
   (\,-\,\delta x_\mu\, P_\mu\, +
      \frac12\,
      \delta\omega_{\mu\nu}\, J_{\mu\nu} \,) \geqslant \hbar \ ,
\]
где $P_\mu$~--- векторный оператор энергии-импульса,
$J_{\mu\nu}$~--- тензорный оператор четырёхмерного момента
импульса, $\delta x_\mu$ и $\delta\omega_{\nu\rho}$~---
стандартные логарифмические координаты группы Пуанкаре.

Что же касается применения принципа неопределённости Гейзенберга,
то оно вряд ли возможно. В предыдущем пункте мы видели, что для
величин время - энергия это вызывало большие трудности.

В связи с этим, уже из соображений релятивистской инвариантности
ясно, что даже для координат и импульсов дело не должно обстоять
просто. И это действительно так, поскольку известно, что все
попытки вводить понятие координат (как самосопряжённых операторов
на пространстве состояний) для релятивистских систем носят
довольно искусственный характер\footnote{Для некоторых же систем
оказывается, что ввести хорошие операторы координат и вовсе
невозможно~\cite{NewtWig1949}.}.

